\documentclass[final,,twoside]{IEEEtranTCOM}

\normalsize

\ifCLASSINFOpdf

\else

\fi

\normalsize
\ifCLASSINFOpdf
\else
\fi

\usepackage{amsthm}
\usepackage{amsmath}
\usepackage{amssymb}
\usepackage{graphicx}
\usepackage{esint}
\usepackage{psfrag}
\usepackage{amsfonts}
\usepackage{cite}
\usepackage{balance}
\usepackage{caption}
\usepackage{subcaption}
\usepackage{epstopdf}
\usepackage{color}
\usepackage{afterpage}
\definecolor{custom_green}{rgb}{0.0, 0.5, 0.0}

\makeatletter

\theoremstyle{plain}

\def\endthebibliography{%
	\def\@noitemerr{\@latex@warning{Empty `thebibliography' environment}}%
	\endlist
}

\makeatother

\providecommand{\theoremname}{Proposition}

\renewenvironment{IEEEbiography}[1]
  {\IEEEbiographynophoto{#1}}
  {\endIEEEbiographynophoto}

\begin{document}
\title{Optical wireless communications for in-body \\ and transdermal biomedical applications
}

\author{Alexandros-Apostolos A. Boulogeorgos (Senior Member, IEEE), \\
Stylianos E. Trevlakis (Student Member, IEEE),  and
Nestor D. Chatzidiamantis (Member, IEEE)

\thanks{The authors are with the Department of Electrical and Computer Engineering in Aristotle University of Thessaloniki, Greece (E-mails:al.boulogeorgos@ieee.org, \{trevlakis, nestoras\}@auth.gr).}
}
\maketitle

\begin{abstract}

This article discusses the fundamental architectures for optical wireless systems for biomedical applications. After summarizing the main applications and reporting their requirements, {we describe the characteristics of the transdermal and in-body optical channels as well as the challenges that they impose in the design of communication systems.} In more detail, we provide three possible architectures for transdemal communications, namely electro-optical (EO) monitoring, opto-electrical (OE) and all-optical (AO) for neural stimulation, which are currently under investigation, whereas for in-body communications, we provide a nano-scale AO (NAO) concept. For each architecture, we discuss the main operation principles, the technology enablers and research directions for their development. Finally, we highlight the necessity of designing an information theoretic framework for the analysis and design of the physical (PHY) and medium access control (MAC) layers, which takes into account the channels'~characteristics. 
\end{abstract}

\section{Introduction}\label{S:Intro}
Medical implants (MIs) and nano-scale wireless networks (NWNs) have been advocated as an effective-solution to numerous health-issues. Typical MIs consist of an out-of-body (in-body) unit that captures the stimulus (bio-signal), converts it into a radio-frequency (RF) signal and wirelessly transmits it into the in-body (out-of-body) unit, which stimulates (monitors) the corresponding nerve (bio-process). The main disadvantage of this approach is its incapability to support high-data-rates required for neural-prosthesis applications, under reasonable transmission-power. Additionally, there is a lack of effectiveness and flexibility in co-existing with the huge amount of RF-devices that are expected to conquer the beyond 5G wireless world. Finally, the use of RF-frequencies prohibit the utilization of nano-scale biomedical devices. 

In view of the advances in optoelectronics and optogenetics, very recent innovative designs have been reported, which, by employing optical wireless communications (OWCs), are able to deliver more compact, reliable, and energy-efficient solutions, while, at the same time, reduce the RF-radiation concerns~\cite{Trevlakis2019}. {The main advantages provided by OWCs are the abundant available bandwidth, no interference from other devices, higher achievable data rates and increased safety for the human body.} As a result, OWCs are expected to be used for both transdermal and in-body links establishment; hence, they will influence the fundamental technology trends in biomedical applications for the next $10$ year and beyond. Due to the transdermal and in-body optical channel particularities, the compactness and energy limitations as well as the directionality of the OWC links, the design and development of OWC-based biomedical applications will have to leverage breakthrough technological concepts. Indicative examples are the co-design of communication and stimulation units, the presentation of digital signal processing (DSP) schemes for the end-to-end link from the device (biological unit) to the biological unit (device), and the utilization of {new EO, OE, AO, and NAO interfaces.} Likewise, new channel and noise models that takes into account the peculiarities of the transdemal and in-body optical medium need to be developed. Building upon which, a novel information theoretic framework is required for the design of energy-efficient PHY and MAC schemes, as well as the development of simultaneous wireless information and power transfer (SWIPT) policies and resource allocation~strategies~\cite{A:WPT_strategies_for_implantable_bioelectronics}. 

Motivated by the above, {the aim of this article is to} present the vision and approach of delivering safe and high {quality-of-experience} (QoE) in MIs and NWNs, identify the critical technology gaps, and the appropriate enables. In more detail, after presenting the biomedical applications, which can be OWC-enabled, together with their critical design requirements, we introduce  three possible architectures for transdemal communications, which are currently under investigation, whereas for in-body communications, we discuss the NAO concept. These architectures are expected to drastically increase the achievable data-rate and significantly reduce the in-body devices energy-consumption. For each architecture, we report the main operation principles, technology enablers as well as critical technology gaps. Moreover, we emphasize the need of developing novel channel models that take into account the transdermal and in-body optical medium particularities as well as deriving a corresponding theoretical framework for the performance assessment and design of the PHY and MAC~schemes. 

\section{Beyond classical biomedical applications}\label{S:BeyondCBA}
\begin{figure}
\centering\includegraphics[width=1\linewidth,trim=0 0 0 0,clip=false]{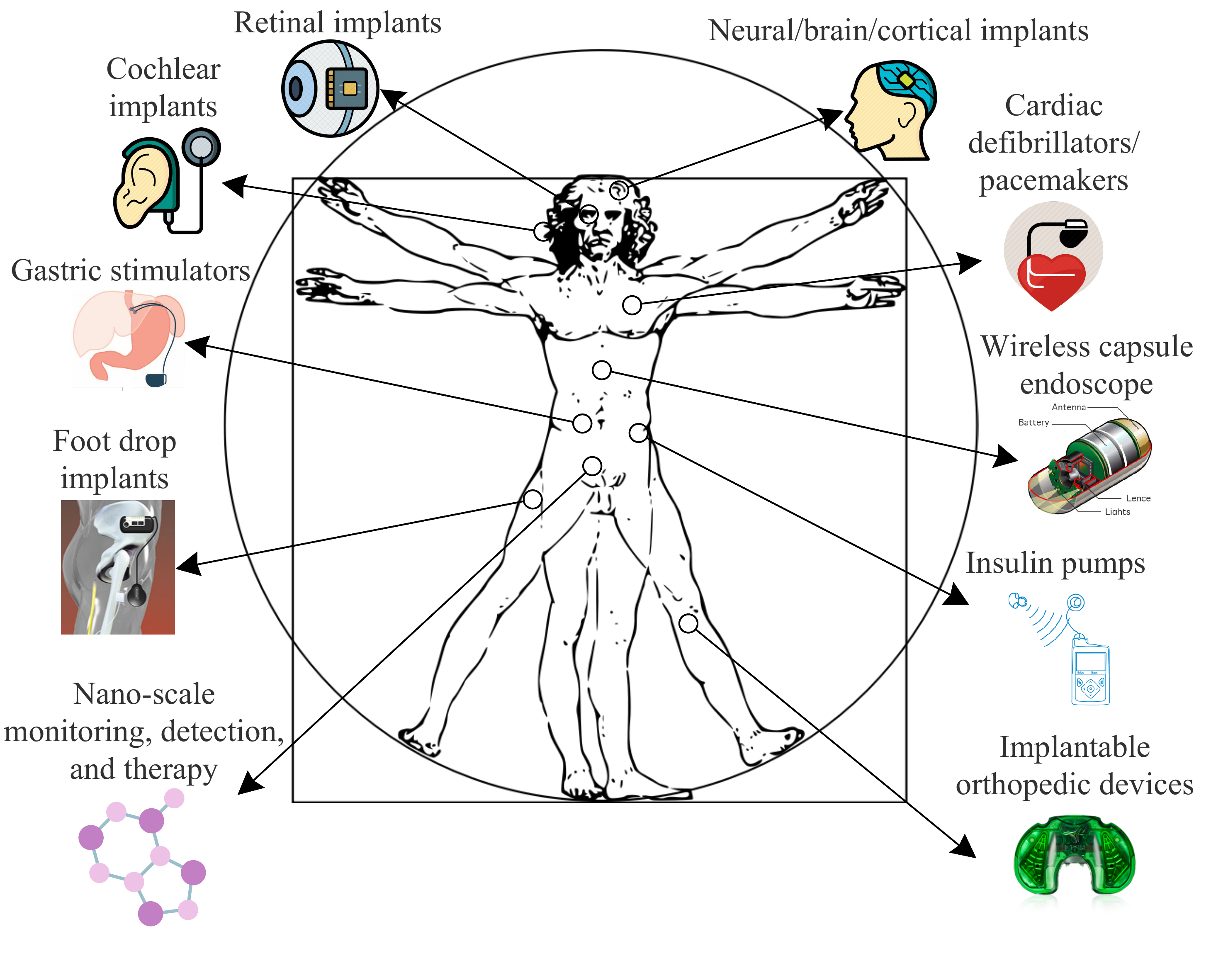}
\caption{Biomedical applications that can be supported by~OWCs.}
\label{fig:AS}
\end{figure}

Optical wireless MIs (OWMIs) and NWNs are envisioned to enable a vast variety of novel biomedical applications, such as smart drug delivery and tissue recovery, as well as to improve the QoE and safety of the conventional ones, not only by targeting $10-100$ times higher-data-rates, but also by combining them with reliability and compact designs. Vital signals, pathogens and allergens real-time continuous monitoring, detection of tissues and molecules abnormalities, as well as smart drug delivery are only some examples of several highly challenging use cases. {In order to support such application scenarios, the research world turns to the adaptation of THz technologies, which, by leveraging graphene designs, is expected to utilize nano-scale transceivers. However, the safety of THz links is still questionable and these technology are yet immature.}  To break these barrier, we need to direct research towards the mature OWC concepts, identify and categorize the possible applications and design suitable architectures and systems~\cite{B:OWC_An_Emerging_Technology}. In this direction, we present the main application categories based on the link nature and scenarios as well as their critical design parameters. These applications are depicted in Fig.~\ref{fig:AS}.

\subsubsection*{Biomedical applications requiring the establishment of transdemal links}  
The main representatives of this category are the  cochlear, retinal, cortical, foot drop, implants, the gastric stimulators, the wireless capsule endoscope, the insulin pumps, as well as the implantable orthopedic devices. Maybe the most successful application of such devices is cochlear implants (CIs), which have restored partial hearing to more than $350,000$ people worldwide, half of whom are pediatric users, who ultimately develop nearly normal language capability. Conventional CIs exploit near-field magnetic communication technologies and typically operate in low-RF frequencies, from $5$ to $49\text{ }\mathrm{MHz}$, while their transmit power is in the order of tens of $\mathrm{mW}$. As a result, they cannot support high-data-rates (in the orders of $\mathrm{Mb/s}$), which are required in order to achieve similar performance to that of cochlea, under reasonable transmit power constraint. Transdermal-optical-link (TOL) is an important building-block to guarantee high-speed-connectivity between the external and internal units with low-energy-consumption~\cite{Trevlakis2018}. In this scenario, apart from the high targeted data-rates $(~10\text{ }\mathrm{Mb/s})$, the critical parameters are the \textit{latency}, which should be in the order of~$0.1\text{ }\mathrm{ms}$, uninterrupted connectivity (i.e., an outage probability that is lower than $10^{-5}$), and low error rate ($\text{BER}\leq 10^{-4}$). Additionally, these devices should support SWIPT in order to guarantee the internal device energy autonomy.  

After the successful clinical validation of CIs, neural prostheses were exploited in the treatment of visual impairments, by developing retinal implants (RIs). In this application, the implanted device is an epiretinal prosthesis that includes a receiver antenna, a DSP unit, and an electrode array attached in to the retina surface. The external unit consists of a miniature-video-camera attached in glasses and a transmitter coil, connected via a small video-processing-unit, which converts the captured-video to electrical-pulse. These pulses are transmitted wirelessly to the implanted device. Unfortunately, the visual acuity of this approach is quite low, due to the limitations of the stimulation unit and the achievable data-rate of the RF-communication-link. To break through this barrier, two solution has been reported, namely, the use of optical stimulation approaches in order to reduce the pixel dimensions and OWCs to increase the date-rate~\cite{Trevlakis2019}. The critical parameters for this application are the pixel population, which should be higher than $60 \times 60$ pixels, the data-rate $(\sim 100\text{ }\mathrm{Mb/s})$, and the latency, which should be less than some decades of ms.

Gastric stimulators are used in the treatment of gastric dysmotility disorders and obesity. They consists of two units, namely out-of-body and in-body. The former's main responsibility is to wirelessly transfer power to the latter, which consists of a received RF-antenna, a charge pump and a peripheral interface controller that generates the electrical-pulse and is connected to the electrodes. The transmission medium is body tissue of $1$ to $5\text{ }\mathrm{cm}$ thickness. This is translated to an about $4\text{ }\mathrm{dB}$ penetration-loss, when RF-signals are employed. However, if optical signals are employed, this loss is further constrained. Except from the transmission range, another critical parameter for this application is the in-body unit compactness.
             
Biomedical implants are also used for bio-signals and bio-processes monitoring, such as neural/brain implants, insulin pumps and wireless capsule endoscopes. Neural/brain implants, pacemakers, and insulin pumps, expect of the high-data-rates, demand high-energy-autonomy and compactness of the in-body unit. In other words, efficient-power-transfer that take into account the space constraints need to be utilized. On the other hand, wireless capsule endoscope is an ad-hoc setup. As a result, its battery can cover the energy demands. The main challenge of this application is to sustain continuous-connectivity of the in-body and out-of-body units independently of the in-body unit orientation.       
    
\subsubsection*{In-body biomedical applications}
Nanotechnology provides a new set of tools to control matter at the atomic and molecular scales, thus, enabling the development of nano-scale biomedical applications for nano-sensing and actuation. Nano-devices utilized in such systems require the use of bio-compatible materials as well as communication techniques with limited electromagnetic radiation on the biological~tissues~\cite{Pisanello2016}. Moreover, several breakthroughs in the field of nano-photonic devices enabled nano-scale PWCs. These applications include, among other, diagnostic and therapeutic techniques utilizing nano-devices, healthcare monitoring solutions that combine in-body nano-photonic bio-sensors and non-intrusive brain-machine interfaces. Among others, nano-LEDs can be leveraged as energy-efficient compact signal sources for optical wireless links, which meet both size and power requirements of nano-devices~\cite{Wirdatmadja2017}. Similarly, sensitive nano-PD were developed and can be combined with nano-LEDs to form nano-sensors, which can be utilized in compact nano-transceivers to forge fast, short-distance wireless communication links. The aforementioned nano-transceivers when paired with nano-antennas, can serve as nano-gateways to bridge the communication with out-of-body devices. 

Despite the achieved progress in the last years, numerous challenges exist that have to be addressed in order to achieve the desired performance of different applications. For example, in the case of wireless network on a chip, the main concern is to increase the capacity and data rate of the network, while dealing with a more predictable and controllable system model. Therefore, new challenges in terms of nano-photonic devices and communication protocols, such as energy-efficiency, link-capacity and reliability, have to be addressed before NAO~networks can be established as a realistic solution. 

\section{Candidate architectures}
\subsubsection*{For transdermal optical links}
\begin{figure}
\centering\includegraphics[width=1\linewidth,trim=0 0 0 0,clip=false]{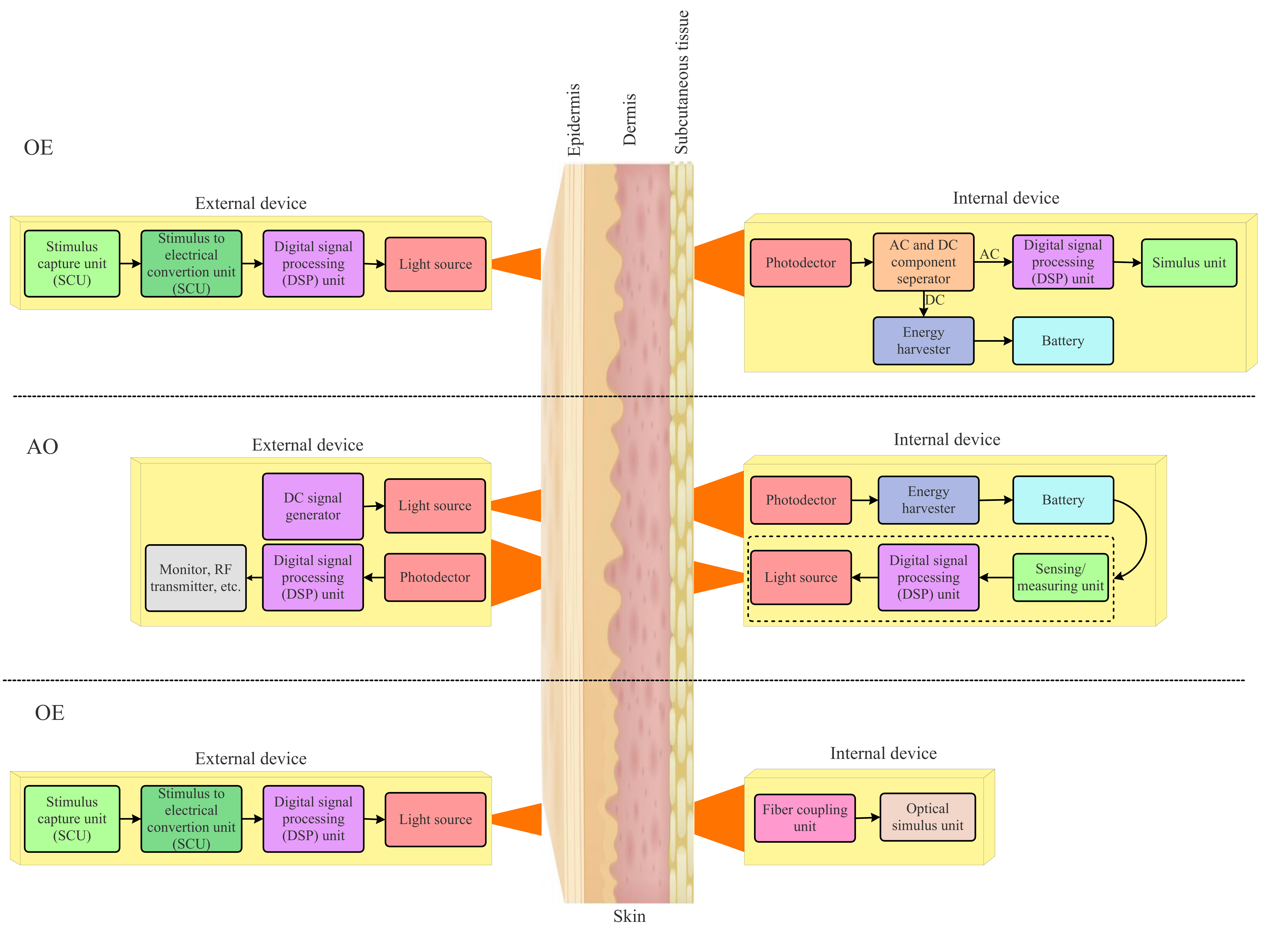}
\caption{Candidate architectures for transdermal optical applications.}
\label{fig:CA}
\end{figure} 

As illustrated in Fig.~\ref{fig:CA}, three candidate architectures for transdermal OWCs have been identified, namely OE, EO, and AO.  OE and AO are suitable for translating external stimuli, such as audio and video, into nerve stimulation signals. Their main difference is that the internal device of the OE unit performs DSP in the received signal. Thus, a energy-autonomy demand for the internal device arises. To satisfy this, a simultaneous light information and power transfer (SLIPT) policy is utilized. In particular, the DC part of the received signal is used for energy-harvesting, while the AC part conveys the neural stimulation message. A simple AC and DC separator (ADS) is employed at the internal device consisting of a capacitor that blocks the DC component of the received signal and forwards it to the energy harvesting branch. At the same time, the AC component of the signal at the output of the ADS is inserted to the DSP-unit, which together with the STM translates the received message into the appropriate neural-stimulation-signal. 

On the other hand, in the AO, no-SLIPT is required, since the stimulation signal is directly emitted by the out-of-body-device. In this architecture, the in-body-device consists of a fiber-coupling-unit, which is usually a MEM and is responsible for the capturing of the transmitted-signal and forwarding it to the optical-stimulator. This concept may be utilized by employing commercially available optics and optogenetics modules. In this scenario, DSP for mitigating the impact of both the TOL impairments and stimulation signal is required. The main challenge of AO is modeling the link from the external-unit to the corresponding nerve. Another challenge is the mapping of the transmission signal into the stimulation one.

The EO architecture can be used for monitoring and sensing bio-signals; hence, it can be employed in neural/brain implants and insulin pumps. The in-body-device consists of a sensing/monitoring-unit, a DSP-unit and a light-source that emits the modulated sensed message. To cover the energy-requirements of the in-body-device an energy-harvesting-branch is utilized. The out-of-body-device consists of a photodetector that captures the light-signal and forwards it in a DSP-unit, responsible for message detection. The detected-signal is forwarded to the imaging-device. Moreover, the out-of-body-unit has a DC signal-generator with a dimming controller  that feeds a light-source, responsible for energy-transfer. The major challenge of EO is to present suitable dimming control strategies that guarantee the energy-autonomy of the in-body-device. Moreover, this concept allows the integration of the internal device front-end into a single-module with superior compactness and energy-efficiency.            

\begin{figure}[t]
\centering\includegraphics[width=0.9\linewidth,trim=0 0 0 0,clip=false]{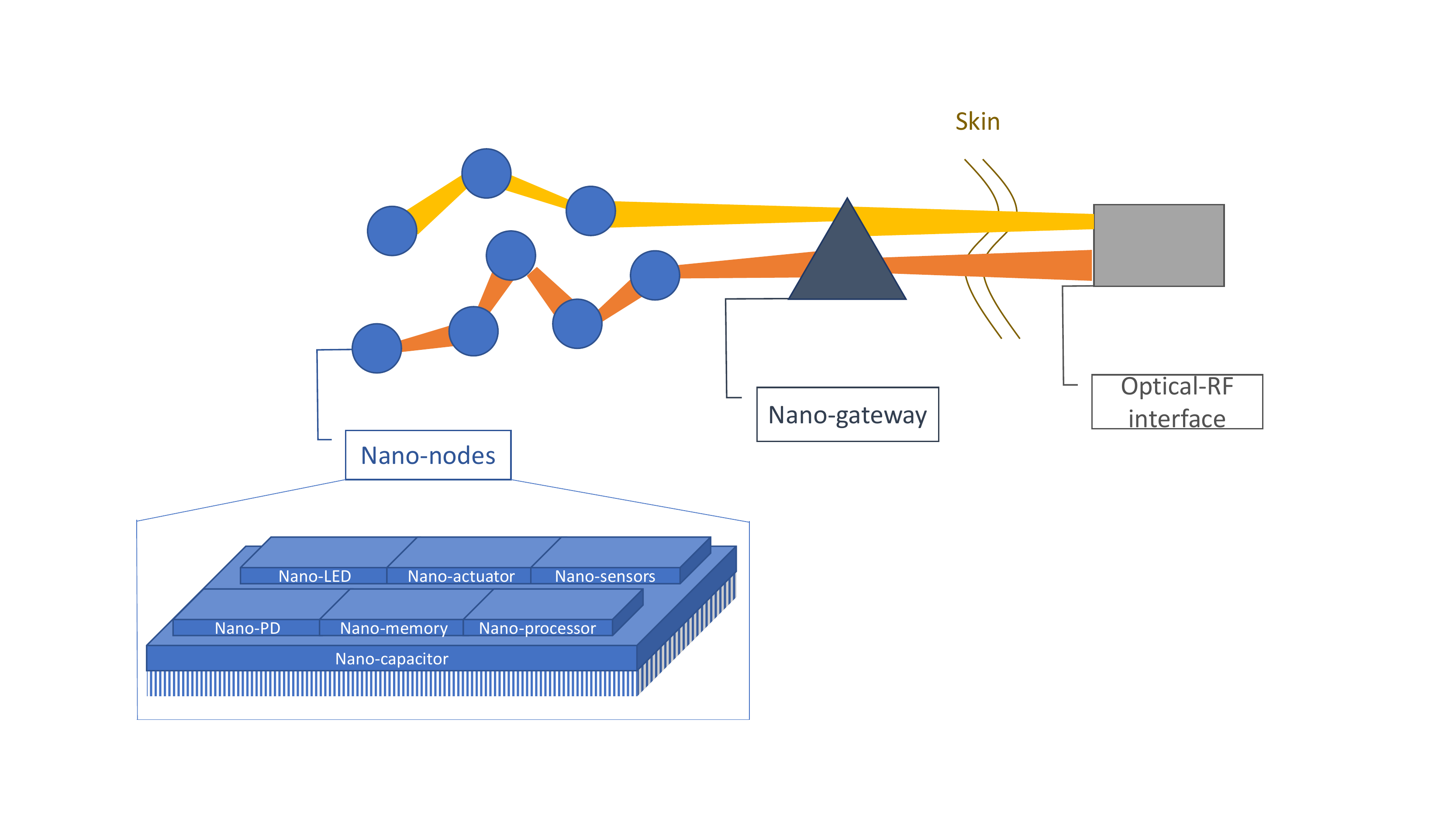}
\caption*{(a)}
\centering\includegraphics[width=0.9\linewidth,trim=0 0 0 0,clip=false]{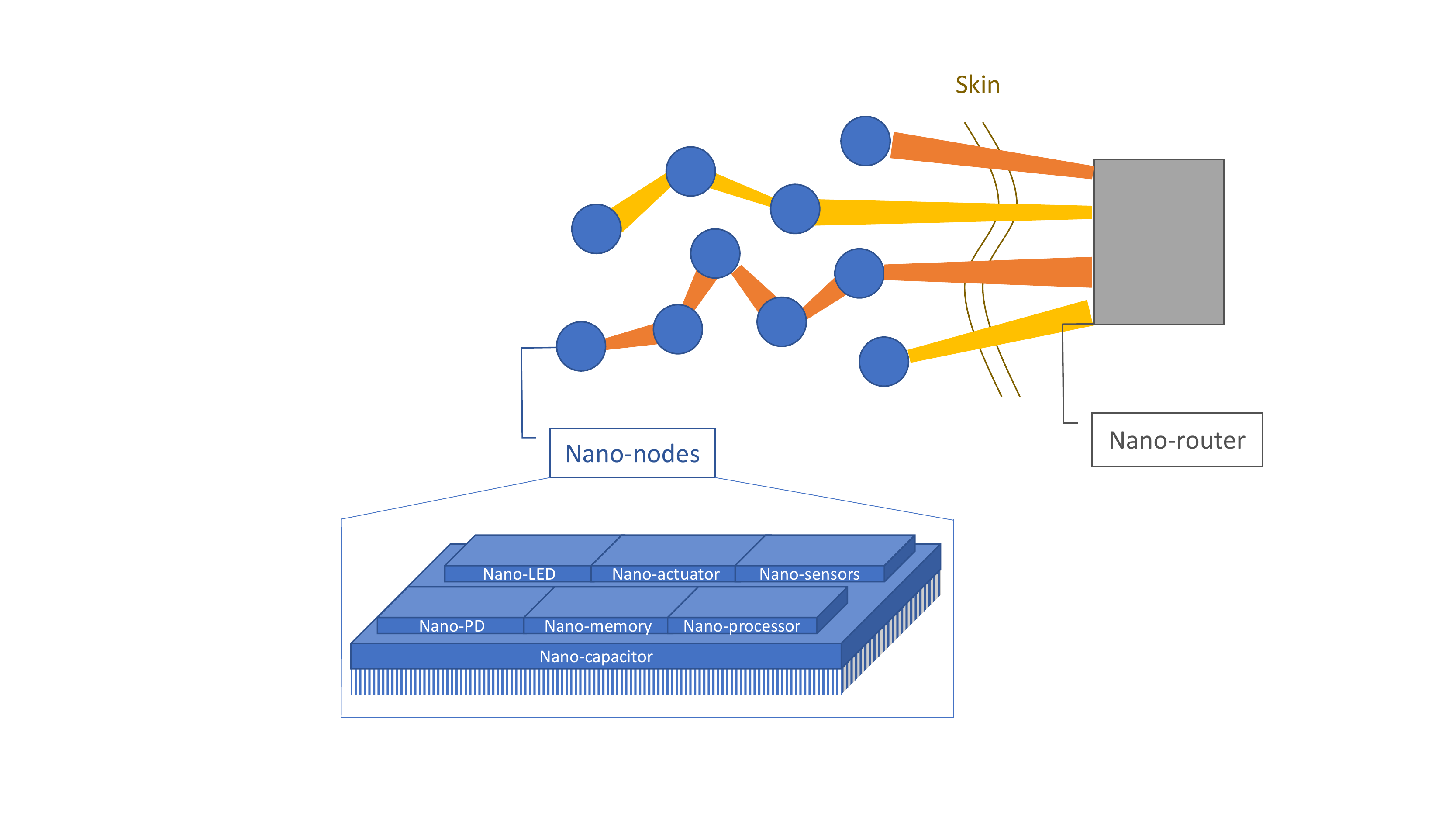}
\caption*{(b)}
\caption{Candidate architectures for in-body optical applications.}
\label{fig:CA_inbody}
\end{figure} 

\subsubsection*{For nano-scale systems}
For the utilization of NAO networks, the two architectures depicted in Fig.~\ref{fig:CA_inbody} have been identified. A common nano-node structure is considered in both cases. Among others, each nano-node is equipped with a nano-LED and a nano-PD that enable communication with other nano-nodes, as well as an energy-harvester that replenishes the energy stored in a nano-capacitor. This enables circulating nano-nodes to overcome the energy-limitations and even have practically infinite lifetime. Depending on the application, the nano-nodes can be equipped with nano-sensors and/or nano-actuators for the monitoring and/or manipulation of biological processes,~respectively. 

Initially, the architecture presented in~Fig.~\ref{fig:CA_inbody}(a) incorporates an intermediate nano-device, called nano-gateway. This device is responsible for forwarding the data collected from the nano-nodes to the external interface. The link between the latter and the nano-gateway can be established on the optical or the RF spectrum, depending on the application characteristics. For example, in close range communications the optical link has been proven to outperform the corresponding RF, while for longer range applications the required optical beam intensity for achieving adequate signal quality can be destructive to the human tissue due to extensive increase in temperature. 

In the scenario depicted in~Fig.~\ref{fig:CA_inbody}(b), the architecture consists of the nano-nodes and the nano-router. The former move throughout the body with a pattern that varies with the application, while the latter collects the gathered information when the nano-nodes are in range~\cite{CanovasCarrasco2018}. For example, if we consider a scenario where the nano-nodes are injected into the bloodstream and the nano-router is placed on the outer surface of the skin, as the nano-node travel through the circulatory system, they communicate with the nano-router when they are positioned close enough to~it.

The main differentiating factor between the two candidate architectures is scalability. Specifically, the aforementioned structure of the nano-network is capable of being multiplied throughout the human-body, in both deep-tissue and near the surface alike. By placing the nano-gateways with appropriate distance between them, the architecture shown in~Fig.~\ref{fig:CA_inbody}(a) can be repeated multiple~times. However, both architectures share the same concerns such as limited energy consumption, limited computational capabilities, stochastic network topology.

\section{Design Principles And Technology Enablers}
The utilization of transdermal and in-body OWC systems exploits the joint-design of information, power-transfer and nerve-stimulation signals that can countermeasure the peculiarities of the optical-channels. Aspired by this, this section is devoted on identifying the channel characteristics and the enabling technologies for the implementation of the candidate~architectures.  

\subsection{Optical wireless channel}
In-body and transdermal optical-wireless channel models are considered to be the main building-blocks for developing the aforementioned architectures. In comparison with the out-of-body models, these ones are required to accommodate a number of different characteristics and biological particularities. For instance, in TOLs, the dermis content in hemoglobin can significantly influence its light absorption in the blue and green-yellow regions. On the other hand, the dominant absorber in the ultraviolet and infrared regions is  the epidermis melanin. Finally, chromophores, like bilirubin, carotene, lipids, cell nuclei, filamentous proteins, etc. can cause further absorption in different wavelengths. Of note, in spite of the abundance in all tissues, due to the short communication distance, water is not a significant light absorber in these systems. Of note, this is one of the differences between the transdermal/in-body and out-of-body optical channels. 

\begin{figure}[t]
	\centering\includegraphics[width=1\linewidth,trim=0 0 0 0,clip=false]{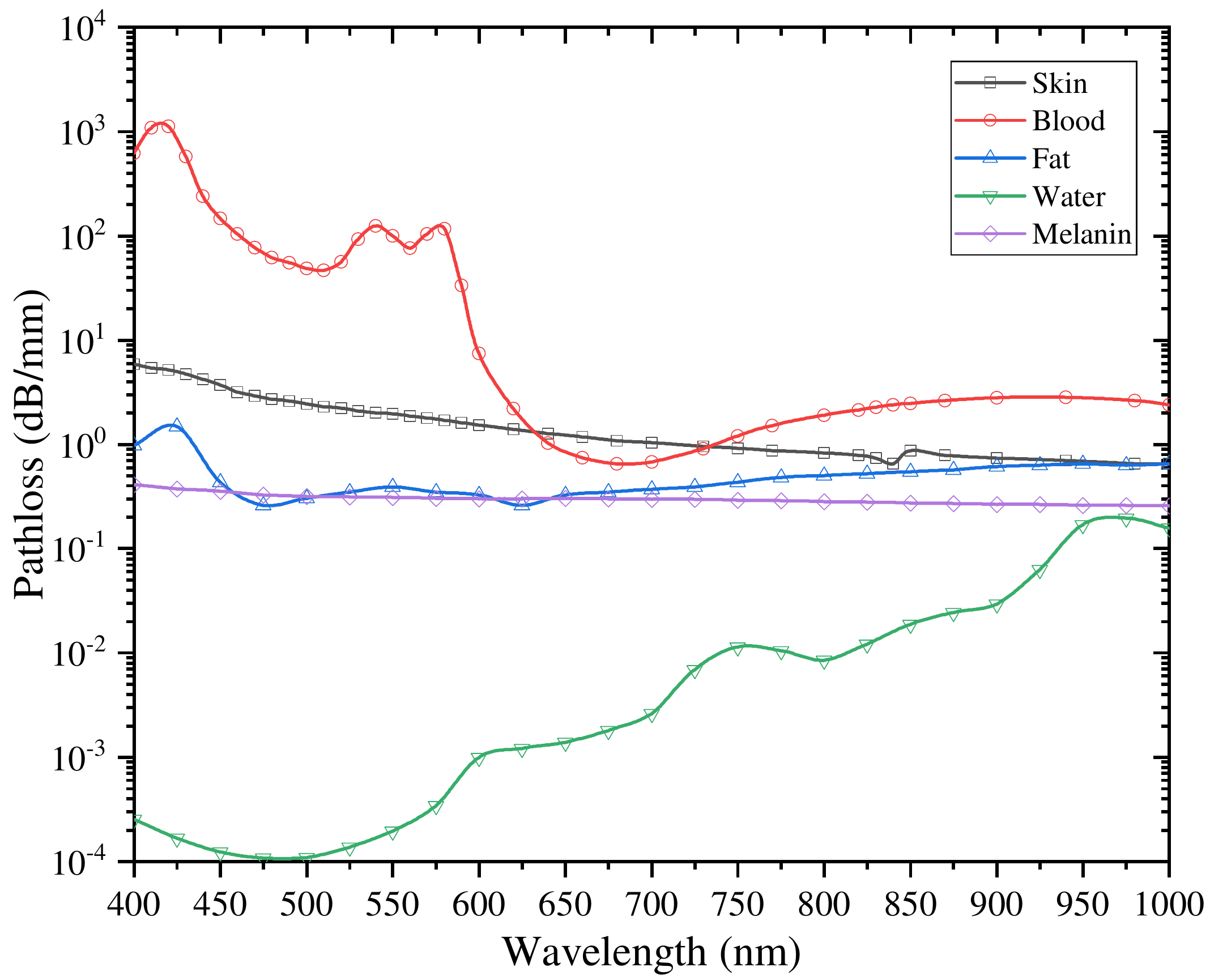}
	\caption{Optical wireless channel path-loss.}
	\label{fig:Pathloss}
\end{figure} 

The impact of the optical wireless channel particularities with regard to path-loss is presented in Fig.~\ref{fig:Pathloss}. This figure summarizes the main contributing elements of any tissue which is applicable to the communication paradigms under investigation. Moreover, the path-loss that occurs when light of a specific wavelength travels through a tissue is a function of the absorption coefficient of the tissue, which can be calculated based on the tissues concentrations and absorption coefficients of water, blood, fat and melanin~\cite{B:Transdermal_optical_communications}. From Fig.~\ref{fig:Pathloss}, we observe that the attenuation due to the existence of blood plays a detrimental role in the performance of the communication link. If we consider only the blood, it is evident that a transmission window exists after $600\text{ }\mathrm{nm}$. On the contrary, water attenuates more intensively optical signals with higher wavelength, while the rest of the elements have a more consistent impact. As a result, the appropriate transmission wavelength for each use case has to be selected based on the composition of the intermediate tissue.

Apart from path-loss, transdermal and in-body optical links experience wavelength-dependent particulate scattering. In particular, within the skin, the main source of scatter is filamentous proteins (like keratin in epidermis and collagen in dermis). Note than since these particles are comparable or larger than the transmission wavelength, scattering can be approximated as a Mie solution to Maxwell's equations. On the other hand, in in-body applications, tissues, such as membranes, striations in collagen fibrils, macromolecules, lysosomes, vesicles, mytochondria, and nuclei, are the main scatters. Notice that membranes are usually lower than $1/10$ of the wavelength, while all the other structures are comparable to the wavelength. Thus, scattering in tissues can be modeled as a mixture of Rayleigh and Mie processes.      

Inhomogeneities in the body content in light absorbing and scattering structures leads to a variation of the reflective index along the transmission path, which cause random fluctuations in both the amplitude and phase of the received signal, i.e. channel fading. In order to provide the theoretical framework for the performance assessment and design of optical biomedical implants, it is of high importance to deliver a stochastic model for the accommodation of this type of fading. In this direction, stochastic geometry approaches have been employed (see~\cite{Guo2016} and references therein). Finally, another source of received power uncertainty, which arises from the directional nature of the optical links, is transceiver misalignment. As reported in~\cite{Trevlakis2018}, this type of uncertainty can be modeled through a stochastic process and can significantly affect the system~performance.

{Another differentiation of in-body/transdermal optical communications in comparison to other out-of-body applications is the existence of neural noise.} Specifically, optical approaches for tracking neural dynamics have modeled the physical bounds on the detection of neural spikes as photon counting statistics (shot noise). Besides the neural noise, other noise sources come from the receiver, like thermal, background and dark current noises, and can be modeled as zero-mean Gaussian processes. Their power depends on the communication bandwidth, the detector's responsivity, the background optical power and the intensity of the dark current, which is generated by the photodetector in the absence of background light. Depending on the architecture, different types of noise determines the fundamental bounds. For example, in AO, neural noise is expected to play the most detrimental role, while, in OE and EO, its intensity tends to zero.

The above observations motivates the development of generalized deterministic path-loss and statistical in-body-particulate models for TOLs and in-body optical links that will aid to the appropriate transmission waveform design, the development of the PHY and MAC, as well as the utilization of the candidate architectures. These models should take into account the patients particularities, such as the skin color, the composition in different structures (collagen, melanin, etc.), the transmission wavelength, and the intensity of misalignment.


\subsection{Electrical and optical neural stimulation}
Based on the stimulation type, implants can be classified into electrical and optical. Electrical stimulation methods use an external electrical current to manipulate the membrane potential directly and have been extensively used for igniting action potentials in various types of human cells. Numerous medical implants, such as cochlear implants, pacemakers, and more, take advantage of these techniques. However, the performance of electrical cell stimulation is hindered by poor spectral coding and bandwidth scarcity of the RF spectrum. These limitations motivated the research community to turn its attention to the optical spectrum, which is capable of providing large amounts of unexploited bandwidth and higher safety for the human body~\cite{Jornet2019}. As a result, optical stimulation has been greatly investigated over the past decade, due to the various advantages it offers over the established electrical one. Optogenetic techniques are based on the development of several genetically modified proteins, called opsins, which when illuminated generate a flow of ions through the cellular membrane that alters the membrane potential~\cite{Ferenczi2019}. This implementation has been experimentally proven to control the cell dynamics with higher accuracy, frequency and greater spatial resolution.

\begin{figure}[t]
	\centering\includegraphics[width=1\linewidth,trim=0 0 0 0,clip=false]{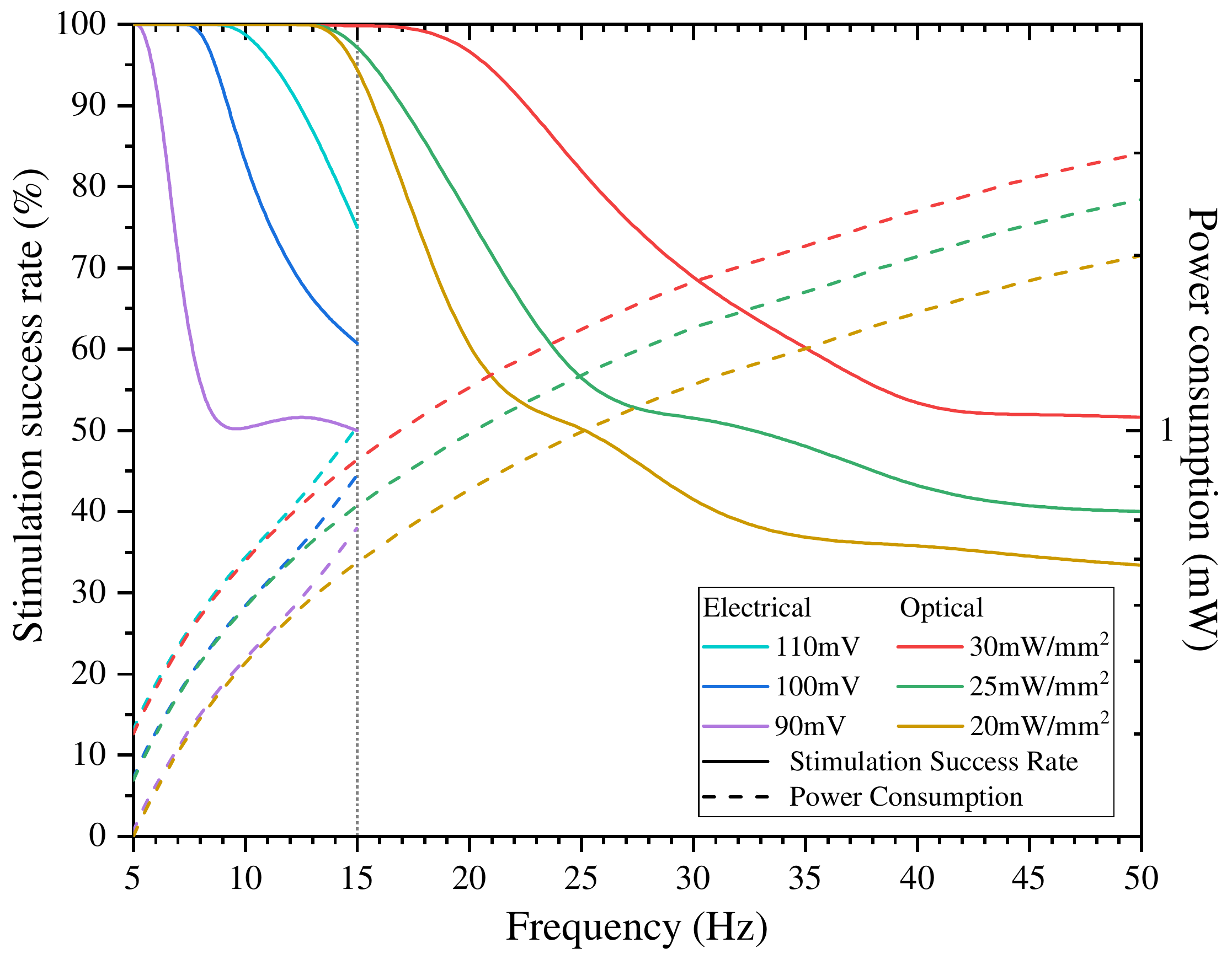}
	\caption{Optical versus electrical neural stimulation performance.}
	\label{fig:Optical_vs_Electrical}
\end{figure} 

Regarding their performance comparison, as indicated~in~Fig.~\ref{fig:Optical_vs_Electrical}, under the same power consumption and stimulation frequency, optical stimulation is proven to outperform the equivalent electrical one in terms of accuracy and power consumption, respectively. It should be noted that the vertical doted line, which corresponds to the $15\text{ }\mathrm{Hz}$, marks the upper frequency limit for the electrical stimulation. It is evident that the optical stimulation is capable of generating accurate action potentials with higher frequency while maintaining the similar power consumption. This observation is in agreement with several publications, which state that optogenetics offer more precise control over the excited neurons due to both the higher achievable frequency and the temporal accuracy provided by the variety of existing photosensitive opsins. In addition, the optical cell stimulation outperforms the corresponding electrical in terms of the stimulation success rate. This is expected due to the fact that the optical spectrum offers more bandwidth. However, if we consider the fact that optical signals attenuate faster as they travel through the human body thus offering lower geometric spread, or in other words, higher spatial resolution, we can deduct that optical stimulation is an overall more robust solution for cell stimulation.

\subsection{Energy transfer and SLIPT}

A critical system design parameters in several inbody biomedical devices is their lifetime, which depends on the amount of available power. Replacing or removing and recharging the battery of such device is impractical or even impossible. To counterbalance this, energy harvesting from the body using thermoelectric, piezoelectric, and electromagnetic generators attracted the eye of biomedical engineers~\cite{Selvarathinam2016}. However, these approaches have two inherent disadvantages: (i) low-harvesting-efficiency ($\sim 20\%$); and (ii) require the installation of extra modules that increase the implementation cost and size. As a consequence,  SLIPT policies seem to be more appealing for biomedical applications. 

In biomedical devices with SLIPT capabilities, the PD, which is a common-choice, due to the high-data-rate that supports, needs to be replaced with solar cells, which provide apparent advantages in terms of harvesting-efficiency~\cite{Wang2015,Wangatia2020}. Solar cells dimensions can be different, depending on the application for which they are used, from some nano-meters to decades-of-centimeters. They are basically photoelectric converters with zero bias and can be used to support both energy-harvesting and information-decoding. From the SLIPT utilization point of view, two fundamental policies can be used, namely time and signal splitting. In the former, the transmission period is divided into an energy-harvesting and information-transfer one, while, in the latter, the DC and AC part of the received photo-current are respectively used for energy-harvesting and information-decoding. 

Although SLIPT approaches and policies have been extensively investigated in different types of OWC systems, they cannot be straightforwardly be applied in biomedical systems, because of the particularities of the receiver. 
In particular, receivers designed for out-of-body OWC applications and have SLIPT capabilities are equipped with solar panels. On the other hand, in biomedical applications, they are equipped with solar cells. As a consequence, further investigation on the suitability of each policy in each one of the aforementioned applications need to be conducted. 

\subsection{PHY \& MAC layer }
By exploiting the capabilities of the optical transceivers, we report the new challenges and opportunities at the PHY and MAC, including energy efficient modulation and coding (MC), energy transfer and SWIPT policies. Consistently with the demands of ultra-low power consumption, compact-size, reliability and ultra-low complexity,  new MC schemes that takes into account the different type of noise and channel particularities need to be developed. Intensity modulation  and direct detection with power adaptation may be a good approach to deal with stochastic behavior of the channel, whereas transmission and/or reception diversity through the exploitation of multiple light sources and photo-detectors can be employed to address the pointing errors.     Moreover, the channel particularities and the transceiver limitations require the development of novel channel codes for in-body/transdermal OWCs. The conventional capacity-approaching channel codes are designed to maximize the data-rate for a given transmit power. However, they demand additional transmission power, which, in low-transmission-range communications is usually comparable to, or even larger than, the transmit power. Another limitation that needs to be taken into consideration is the decoding time, which needs to be constrained according to the application requirements.  To break these barriers, we need to characterize the error sources, i.e. both the noise nature and the channel stochastic behavior, and  examine the trade-off between transmission power and decoding time in order to design channel codes that minimize the aggregates transmit and decoding power. A possible approach is the use of low-weight coding schemes that prevent channel errors. In this direction, efficient algorithms to determine the optimal  coding-weight and generate low-weight codes need to be investigated. 

From the MAC point of view, the directional nature of OWCs combined with the high-bit-rates increase the spatial-synchronization requirements in nano-scale optical networks and call for novel MAC protocols that would be able to guarantee transceivers alignment, device fast-tracking and routing. In this direction, a RX-initiated schemes that allow the devices to periodically operate in sleeping,  receiving and transmitting mode may be the solution. During the reception phase, each device locates its neighbors  and announces its availability. When a device identifies that the intended RX or a candidate relay is available, it switches to transmitting mode.  Another functionality that needs to be utilized is fast steering. Optical tracking and steering is a challenge, especially when the requirement is for low-cost, fast, lightweight control systems. Further research will need to address the development of appropriate field-of-view restrictions and effective beam steering and adaptive control systems for nano-scale biomedical applications. Finally, by exploring the opportunities of multi-hop communications, we can drastically reduce the energy consumption. To achieve this, new routing protocols need to be utilized that take into account the energy availability of each nano-node as well as the application delay constraint. 

\section{Conclusions}
In this article, we presented the concept of optical in-body and transdermal optical communications. In more detail, after presenting the possible biomedical applications and identifying their requirements, we described candidate system architectures and features. Due to the fact that their designs are in early stage, it is difficult to determine their final form. However, their enablers and technology modules were reported, namely channel modeling, the development of novel PHY, MAC and energy transfer schemes, and the design of appropriate stimulation units. Finally, future research directions were highlighted.

\section*{Acknowledgment}
This research has been co‐financed by the European Union and Greek national funds
through the Operational Program Human Resource Development, Education and Lifelong Learning, under the call Supporting researchers with an emphasis on young researchers (Part B) (Project code:~99022).

\balance
\bibliographystyle{IEEEtran}
\bibliography{IEEEabrv,refs}

\vspace{-1cm}
\begin{IEEEbiography}{Alexandros--Apostolos A. Boulogeorgos} (S'11, M'16, SM'19)  received the Electrical and Computer Engineering diploma and PhD degrees both from the Aristotle University of Thessaloniki (AUTh) in 2012 and 2016, respectively. From March to August 2016, he was a postdoctoral researcher in the Centre for Research and Technology Hellas, while, from October 2016 and 2019, he works as a postdoctoral researcher in the University of Piraeus and AUTh, respectively.     
\end{IEEEbiography}
\vspace{-1cm}

\begin{IEEEbiography}{Stylianos E. Trevlakis} received the Electrical and Computer Engineering diploma from the Aristotle University of Thessaloniki (AUTh) in 2016. From May to September of 2017 he was a research assistant in the Centre for Research and Technology Hellas, while, since October 2017, he is a PhD candidate and a part of the Wireless Communications System Group at the AUTh.
\end{IEEEbiography}

\vspace{-1cm}
\begin{IEEEbiography}{Nestor D. Chatzidiamantis} (S’08, M’14) was born in Los Angeles, CA, USA, in 1981. He received the Diploma degree (5 years) in electrical and computer engineering (ECE) from the Aristotle University of Thessaloniki (AUTH), Greece, in 2005, the M.Sc. degree in telecommunication networks and software from the University of Surrey, U.K., in 2006, and the Ph.D. degree from the ECE Department, AUTH, in 2012. Since 2018, he has been Assistant Professor at the ECE Department of AUTH. 
\end{IEEEbiography}

\end{document}